\documentclass{aa}
\usepackage{graphics}
\newcommand{\cmg}{cm$^{2}$\,g$^{-1}$}

\newcommand{\um}{$\mu$m}                                 
\newcommand{\mearth}{M$_{\oplus}$}

\newcommand{\gapprox}{$\stackrel {>}{_{\sim}}$}   

\newcommand{\about}{$\sim$}                       
\newcommand{\bpic}{$\beta \, {\rm Pic}$}          

\newcommand{\amin}{$^{\prime}$}                   
\newcommand{\asec}{$^{\prime \prime}$}
\newcommand{\adeg}{$^{\circ}$}

\newcommand{\radot}[4]{\mbox{#1$^{\rm h}$#2$^{\rm m}$#3$\stackrel{\rm s}
{_{\bf\cdot}}$#4}}
\newcommand{\decdms}[3]{\mbox{#1$^{\circ}$#2$^{\prime}$#3$^{\prime \prime}$}}

\newcommand{\adegdot}[2]{\mbox{#1$\stackrel {\circ}{_{\bf \cdot}}$#2}}

\newcommand{\asecdot}[2]{\mbox{#1$\stackrel {\prime \prime}{_{\bf \cdot}}$#2}}

\begin{document}
   \title{
   The 1.2\,mm image of the \bpic toris disk \thanks{Based on observations collected 
   	with the Swedish ESO Submillimeter Telescope, SEST, in La Silla, Chile.}
   	}


   \author{Ren\'e Liseau\inst{1}        \and
           Alexis Brandeker\inst{1}     \and
           Malcolm Fridlund\inst{2}     \and
           G\"oran Olofsson\inst{1}     \and
           Taku Takeuchi\inst{3}        \and
           Pawel Artymowicz\inst{1}
           }

   \offprints{R. Liseau}

   \institute{Stockholm Observatory, SCFAB, Roslagstullsbacken 21, SE-106 91 Stockholm, Sweden \\
              \email{rene@astro.su.se, alexis@astro.su.se, olofsson@astro.su.se, pawel@astro.su.se}
   \and            
              ESTEC/ESA, P.O. Box 299, NL-2200AG Noordwijk, The Netherlands \\
 	      \email{malcolm.fridlund@esa.int}
   \and
              Lick Observatory, University of California Santa Cruz, CA 95064 \\
              \email{taku@ucolick.org}	      
              }

   \date{Received date: \hspace{5cm}Accepted date:}

   \abstract{We present millimeter imaging observations in the 1200\,\um\ continuum
   of the disk around \bpic. With the 25\asec\ beam, the \bpic\ disk is 
   unresolved perpendicularly to the disk plane ($\le 10$\asec), but slightly resolved
   in the northeast-southwest direction (26\asec). Peak emission is observed at the stellar
   position. A secondary maximum is found 1000\,AU along the disk plane in the southwest,
   which does not positionally coincide with a similar feature reported earlier at 850\,\um.
   Arguments are presented which could be seen in support of the reality of these features. 
   The observed submm/mm emission is consistent with thermal emission from dust grains, which are 
   significantly larger than those generally found in the interstellar medium, including 
   mm-size particles, and thus more reminiscent of the dust observed in protostellar disks. 
   Modelling the observed scattered light in the visible and the emission in the submm/mm
   provides evidence for the particles dominating the scattering in the visible/NIR and 
   those primarily responsible for the thermal emission at longer wavelengths belonging 
   to different populations.
            }

   \maketitle

%

\section{Introduction}

Since its discovery by IRAS (e.g., \cite{aumann84}), the \bpic toris system has presented 
the prime example of a dusty disk around a main sequence star, partly
because of its high degree of `dustiness' ($L_{\rm IR}/L_{\star} = 2.5\times 10^{-3}$, 
e.g. \cite{lagrange2000}) and partly because of its relatively close distance to the Earth
(19.3\,pc, \cite{crifo97}), which makes it possible to obtain high quality data over 
the entire spectral range. Recent papers reviewing the physics of 
the disk around \bpic\ include those of \cite{pawel2000}, \cite{lagrange2000} and \cite{zucker2001}.
Considerable uncertainty existed regarding the age of the system, but most recent estimates place 
the stellar age close to only ten million years ($12^{+8}_{-4}$\,Myr, \cite{zuckeretal}). This
could open up the possibility that planet formation (nearing its final phases?) might actually become observable.

Since these reviews were written, new relevant information has been added to our knowledge of
the \bpic\ system: \cite{olof2001} reported the discovery of widespread atomic gas in the 
disk, a result recently confirmed and extended by \cite{alexis2002}. These observations revealed
the sense of disk rotation and that the northeast (NE) part
of the gaseous disk is extending to the limit of the observations by Brandeker et al., viz. to at 
least 17\asec\ (330\,AU) from the star. Several difficulties were encountered with these discoveries,
such as the observed fact that the gas stays on (quasi-)Keplerian orbits, although 
radiation pressure forces in the resonance lines should accelerate the gas to high velocities 
and remove it on time scales comparable to the orbital period. 

This needs to be addressed in the context of
the origin and evolution of the gas and dust, i.e. whether one or both components 
are presently produced in situ in the disk or whether they (at least to some degree)
constitute `left-overs' from the star formation process, being of primordial origin. 
\cite{taku} have considered the interaction of gas and dust in a
circumstellar disk, the dynamical evolution of which is critically dependent on the relative
abundance of these species (see also \cite{lecav98}). Possible observational consequences, 
even relatively far from the central star, may become assessible with modern mm/submm cameras. 
With the aim to compare and to extend the results obtained at 850\,\um\ with SCUBA by 
Holland et al. 1998, we performed imaging observations at longer wavelengths and, in this paper, 
we present the image of the \bpic\ disk at 1200\,\um\ (1.2\,mm).

The SIMBA observations and the reduction of the data are presented in Sect.\,2. Our basic result,
i.e. the 1.2\,mm image of \bpic\ and its circumstellar disk, is found in Sect.\,3, and in Sect.\,4
the possible implications of these observations are discussed, where also other data are consulted. 
Finally, in Sect.\,5, we briefly summarise our main conclusions from this work.

\begin{figure}
  \resizebox{\hsize}{!}{\rotatebox{270}{\includegraphics{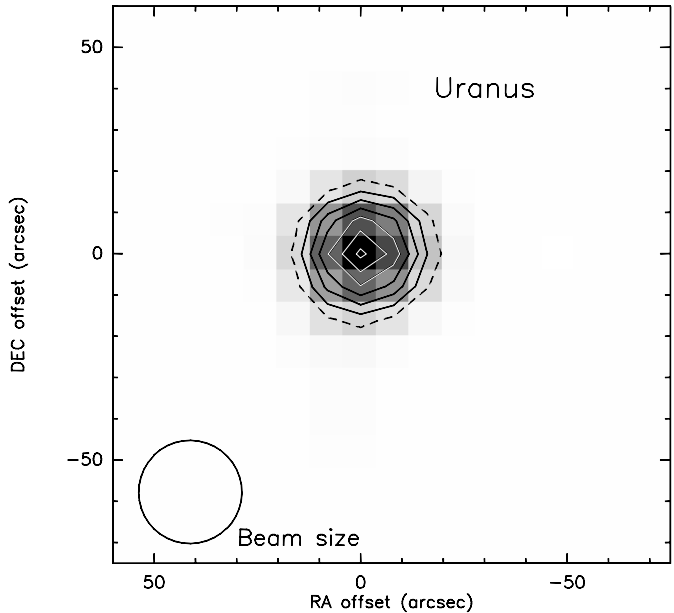}}}
  \caption{The normalised SIMBA 1.2\,mm image with 8\asec\ pixels of the flux calibrator Uranus, 
  the size of which was \asecdot{3}{5} in diameter at the time of our observations and, hence, 
  appeared point-like to SIMBA. 
  Offsets in Right Ascension and Declination are in arcsec and the derived circular Gaussian beam of 
  25\asec\ FWHM is shown in the lower left corner. During the scanning alt-az observations, the
  image rotates which would smear out any low-level features. Contour levels as in Fig.\,\ref{bpic_disk}.
    }
  \label{uranus_psf}
\end{figure}

\begin{figure}
  \resizebox{\hsize}{!}{\rotatebox{270}{\includegraphics{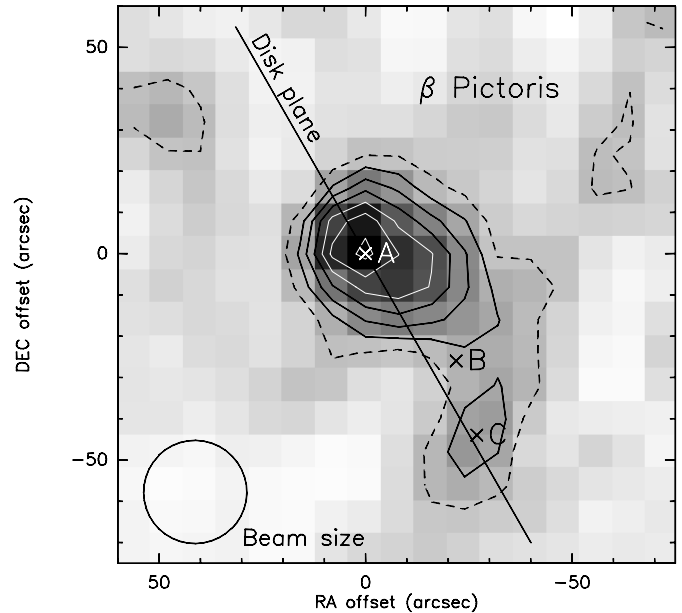}}}
  \caption{The \bpic\ disk imaged at 1.2\,mm with a pixel sampling of 8\asec. The 
  lowest (dashed) contour corresponds to $2\sigma$ and increments are by $1 \sigma =3$\,mJy/beam. 
  The center coordinates (0,\,0) refer to the stellar position and offsets are in arcsec. Features
  discussed in the text are marked and the optical disk midplane is shown by the straight line
  at position angle \adegdot{31}{5}.
  The circular Gaussian beam of 25\asec\ FWHM (\about\,500\,AU) is shown to scale in the lower left corner.
    }
  \label{bpic_disk}
\end{figure}

\section{SIMBA observations and data reductions}

The observations were performed with the SEST Imaging Bolometer Array (SIMBA) at the 15\,m
Swedish-ESO Submillimeter Telescope (SEST), La Silla, Chile, during the periods November 17 - 23
and 27 - 29, 2001. SIMBA has been developed by the Max-Planck-Institut f\"ur Radioastronomie, Bonn, 
in collaboration with the Astronomisches Institut der Ruhr-Universit\"at Bochum.
The 37 liquid helium cooled semiconductor elements are n-doped silicon chips, mounted on a 
saphire substrate. Projected on the sky, the 37 horn antennae have each an HPBW of 23\asec\ and 
are spaced by 44\asec\ in a hexagonal arrangement, covering about 4\amin. The spectral band 
pass is centered at 250\,GHz (1200\,\um) and has a full width at half maximum (FWHM) of 90\,GHz.

As map centre, the equatorial coordinates of \bpic\ were used, viz. 
\radot{05}{47}{17}{1}, $-$\decdms{51}{03}{59} (J\,2000). Generally,
the mapping was done in fast mode by scanning 600\asec\ in azimuth and 400\asec\ in elevation 
at the rate of 80\asec\ per second and with a step size in elevation of 8\asec, oversampling 
the beam by about a factor of three. The zenith opacities were obtained by means of frequent
sky dips and were on the average 0.22 at the beginning of our observing run, but improved 
to about 0.15 after a couple of days. The pointing of the telescope was regularly checked using
the extragalactic radio source $0537-441$ and/or the planet Saturn and was found to be within a
third of the beamwidth. 

Flux calibrations were based on observations of the planet Uranus. The brightness temperature of 
94.8\,K at 250\,GHz implies a Uranus flux density of 38\,Jy/beam, with an 
uncertainty of 5\% (\cite{griffin93}). As is evident from Fig.\,\ref{uranus_psf}, 
the spatial flux distribution of this point source 
(diameter \asecdot{3}{5}) is consistent with a circular Gaussian telescope beam, with the measured 
FWHM($\,\parallel,\,\perp ) = ($\asecdot{24}{9}\,$\pm$\,\asecdot{0}{3}, \asecdot{24}{6}\,$\pm$\,\asecdot{0}{3}).
The deviation from a circularly symmetric Gaussian beam pattern occurs at the 1\% level of the peak value
($-20$\,db).

In total, we recorded 171 maps with 21 hours of integration.
These data were reduced by making use of the MOPSI reduction software
package\footnote{http://www.ls.eso.org/lasilla/Telescopes/SEST/SEST.html} 
developed by Robert Zylka. This involved correcting
for atmospheric extinction, cosmic rays and the variable sky background,
as well as producing maps from the fast scanning mode. The sky noise
was greatly reduced by correlating and removing the simultaneous flux
level variations of the different bolometer channels.

\section{Results}

The 1200\,\um\ image of the \bpic\ region is displayed in Fig.\,\ref{bpic_disk}, where the 
displayed contours are chosen in compliance with \cite{holland98}. The flux maximum,
designated as feature {\it A} in the figure, is centered on the position of the star \bpic\ (0\asec, 0\asec). 
At position angle 237\adeg\ relative to {\it A}, an elongation of the emission in the NE-SW direction is 
discernable. For an assumed Gaussian source flux distribution, the deconvolved minor 
and major axes are $\le 10$\asec\ and 26\asec, respectively, the latter corresponding to 500\,AU.
The disk emission extends to at least 55\asec\ (1050\,AU) in the SW direction. In addition, 
faint emission protrudes south toward a `blob' at ($-27$\asec, $-44$\asec). This feature {\it C}, 
at position angle \adegdot{211}{5}, is 52\asec\ (1000\,AU) distant from the star and is thus not 
(whether real or not) positionally coincident with the 850\,\um-SW blob of \cite{holland98} at 
($-21$\asec, $-26$\asec) and identified as {\it B} in Fig.\,\ref{bpic_disk}, but we note the following important facts: 
(1) this faint emission, extending straight south, is also apparent in the 850\,\um\ image, 
(2) as is evident in Fig.\,\ref{bpic_disk}, the position angle to blob {\it C} coincides with that of 
the midplane of the \bpic\ disk inferred from optical data (= \adegdot{31}{5} + 180\adeg; 
\cite{kalas95}, \cite{heap2000}), and (3) faint dust scattered light in this
direction has been observed far from the star (1450\,AU, \cite{larwood2001}). Toward the NE,
the scattering disk has been claimed to extend even further from the star, to 1835\,AU. 

An about $2\sigma$ feature at ($+48$\asec, $+32$\asec) and pa\,=\,57\adeg\ (1100\,AU)
is discernable in our image, but which would again be significantly offset from a similar blob 
at ($+28$\asec, $+25$\asec) in the 850\,\um\ data. In order to assess the reality of these faint features
we have divided the raw data into different portions and then applied the same reduction procedures 
to these subsamples. The result of this exercise indicates that the ($+48$\asec, $+32$\asec) feature 
is an artefact introduced by the noise, as it is not seen in all frames, whereas blob {\it C} is persistently 
present in the final sub-maps.
 
The flux densities of both the \bpic\ disk ({\it A}) and the SW features ({\it B} and {\it C})
are presented in Table\,\ref{flux_tab}. The result for the peak flux obtained by \cite{chini91}\footnote{We 
assume that these data refer to \bpic\ and not to $\alpha$\,Pic as written in their Table\,1.} 
with a single element bolometer at the SEST at 1300\,\um, viz. $24.9 \pm 2.6$\,mJy/beam, is 
in agreement with our array value of $24.3 \pm 3.0$\,mJy/beam at 1200\um\ (cf. Fig.\,\ref{bpic_SED}). 
The ratio of the peak flux {\it A} to that of blob {\it C} is $2.5 \pm 0.8$. 

\begin{table}
  \caption{\label{flux_tab} SIMBA 1200\,\um\ flux densities of the \bpic\ disk}
 \resizebox{\hsize}{!}{
  \begin{tabular}{clcl}
    \hline
Feature & Relative          & $F_{\nu}(1200$\,\um)  & Remarks  \\
        & Offset (\asec)    & (mJy/beam)            &           \\
    \hline 
\\
{\it A} &(0, 0)         &  $ 24.3  \pm 3.0  $   & {\bf $\beta$\,Pic disk}  \\
        &               &  $ 35.9  \pm 9.7  $   & integrated over a radius of 40\asec   \\ 
\\
{\it B} &(-21, -26)     &  $ \ldots  $          & {\bf SW blob} (\cite{holland98}): \\
        &               &                       & contaminated by \bpic\ disk \\    
\\
{\it C} &(-27, -44)     &  \phantom{1}$ 9.7\pm 3.0$   & {\bf SW blob} (this paper)   \\
\\
 \hline
  \end{tabular}
                       }
\end{table}

\section{Discussion}

\subsection{The \bpic\ disk}

For widely adopted parameters, the stellar disk subtends an angle in the sky of less than 1\,mas and the
photosphere generates a flux density at the Earth of less than 1\,mJy at 1200\,\um. The stellar contribution
to our SEST measurements can therefore be safely ignored. Also, any line emission in this band pass is 
likely to be totally negligible (\cite{liseauarty98}, \cite{liseau99} and references therein). 

Assuming an opacity law of the form $\kappa_{\nu}= \kappa_0\,(\nu/\nu_0)^{\beta}$, the flux density 
at long wavelenghts from an optically thin source of a certain dust population can be expressed as 

\begin{equation}
F_{\nu} = \frac {2 k \kappa_0 \Omega}{c^2 \nu^{\beta}_0}\,\nu^{2+\beta} \int\! T_{\rm dust}(z)\,{\rm d}z\,\,\, .
\end{equation}

\begin{figure}
  \resizebox{\hsize}{!}{\rotatebox{00}{\includegraphics{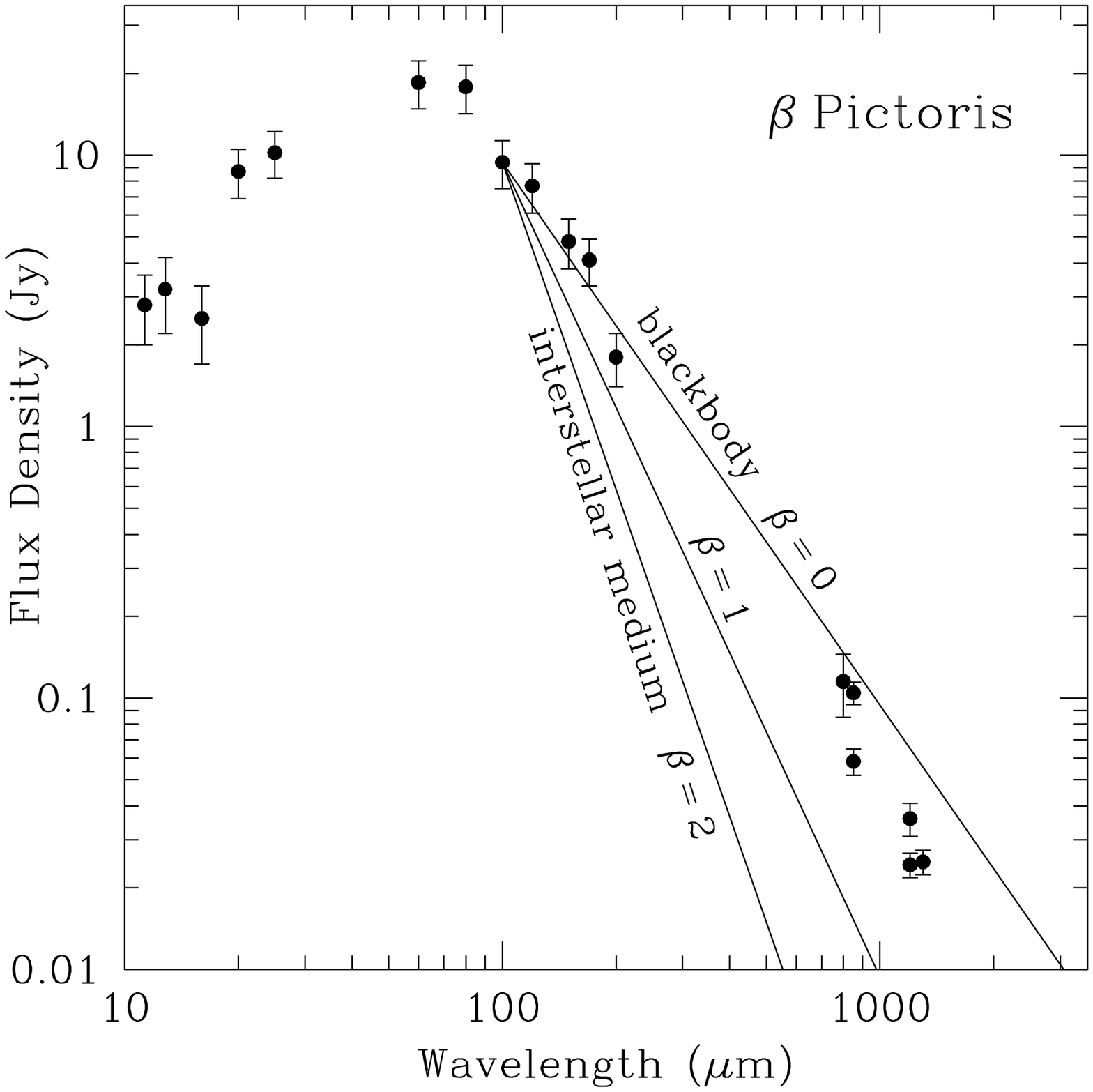}}}
  \caption{The SED of the \bpic\ disk from 10 to 1300\,\um. Data up to 200\,\um\ are from 
  \cite{heinrichsen99} (52\asec\ to 180\asec). 
  The datum at 800\,\um\ is from \cite{zuckerbeck93} (integrated over a radius = 25\asec).
  The 850\,\um\ data are from \cite{holland98} and refer to a 14\asec\ beam and to the integrated flux over a
  radius = 40\asec. The latter should be comparable to our 1200\,\um\ point for 40\asec; the lower is for a
  25\asec\ beam, and that at 1300\,\um\ (24\asec\ beam) is from \cite{chini91}. For reference, 
  the spectral slopes longward of 100\,\um\ for three values of the dust emissivity index $\beta$, 
  discussed in the text, are shown by the straight lines.
    }
  \label{bpic_SED}
\end{figure}

If the particles dominating the emission at 850\,\um\ and 1200\,\um, respectively, can be
assumed to yield the same integral, e.g. because they
share the common temperature $T_{\rm dust}$ and/or occupy similar locations in 
space along the line of sight $z$, the average spectral index $\beta$ can be obtained 
from our 1200\,\um\ and the 850\,\um\ fluxes by \cite{holland98}, using

\begin{equation}
\beta = - 2 + \frac{ {\rm d}\log {F_{\nu}} }{ {\rm d}\log {\nu} } =
-2 + \frac { \log { (F_{850}/F_{1200}) } }{ \log {(1200/850) } }\,\,\, ,
\end{equation}

yielding $\beta = 0.5$ for a point source at the stellar location. The index becomes
$\beta = 1$, if we use the fluxes for the extended source, viz. integrated over a 
radius of 40\asec\ centered on the star. This includes blob {\it B}, contributing some
20\% to the 850\,\um\ flux.  A `correction' for this would again indicate a lower $\beta$ value
and we conclude that the dust in the \bpic\ disk has a shallow opacity index, perhaps even below 
unity (\cite{dent2000} suggest $\beta = 0.8$). This is illustrated in Fig.\,\ref{bpic_SED}, which displays the long wavelength spectral energy
distribution of \bpic, together with weighted Rayleigh-Jeans spectra for $\beta= 0,\,1\,{\rm and}\,2$, 
and could mean that the grains dominating the millimeter-wave emission are different from
those scattering most efficiently in the visual and the near infrared.
This value of $\beta$ is significantly lower than those found in the interstellar medium (ISM), 
where typically $\beta \sim 2$ (\cite{hildebrand83}), but it is similar to that found in {\it protostellar}
disks (e.g., \cite{Beckwith90}, \cite{Mannings94}, \cite{Dutrey96}),
indicating significant differences between the dust particles in the ISM and those in the \bpic\ disk. 
As was also already concluded by \cite{chini91}, the presence of relatively large grains is suggested, 
with maximum radii in excess of 1\,mm ($\max{a}/\lambda$\,\gapprox\,1). Intriguing, however, is the existence of 
such large grains possibly as far away as 1000\,AU from the star (see, e.g., \cite{taku}, \cite{lecav98}).

\subsection{Weak dust features in the 1.2\,mm image}

The asymmetric flux distribution displayed in Fig.\,\ref{bpic_disk} may be surprising. In agreement with
our observations, \cite{chini91} and \cite{dent2000} too were unable to detect any emission in the northeast part of the disk,
where we place a $3 \sigma$-upper limit on the mass of 0.2\,\mearth\ (see below; the dust temperature at 500\,AU
$T_{\rm dust}=45$\,K, when $T_{\rm dust}(r) = 110\,(r/26\,{\rm AU})^{-0.3}$\,K, see \cite{liseauarty98}). 

Given the low signal-to-noise ratio (S/N), the reality of this lopsidedness is difficult to assess, but
asymmetries in the \bpic\ disk have been noticed also at other wavelengths. For instance, in scattered 
light, the receding NE side of the disk extends much further and is much brighter than the SE disk. 
In contrast, the shorter, approaching SE disk seems much thicker (\cite{kalas95}, \cite{larwood2001}). The
situation is reversed in the thermal infrared (albeit on smaller spatial scales), where the SW disk 
appears significantly brighter and more extended than the NE side (\cite{lagage94}, \cite{wahhaj02}, 
\cite{weinberger02}). This could be due to a `Janus-effect', i.e. the NE being dominated 
by `bright' dust particles (high albedo, silicates), whereas in the SE, the majority of dust grains is `dark'
(high absorptivity, carbonaceous?). What would accomplish such uneven distribution in the disk
is not clear, but large differences in albedo, by more than one order of magnitude, are not uncommon,
for instance, in solar system material. Also, in order to understand the nature of feature {\it C} 
(and {\it B}) velocity information would be valuable.

Blob {\it C} would be situated in the disk midplane and on the {\it second} contour in the scattered light 
image of \cite{larwood2001} ($22 < R < 25$\,mag/arcsec$^2$, see also \cite{kalas95}). No obvious 
distinct feature is seen at its position. However, to be detectable with SIMBA at the SEST, 
any point source at mm-wavelengths would not be point-like at visual wavelengths, and its optical 
surface brightness could be very low. 

To gain some
quantitative insight we ran numerical models, exploiting Mie-theory, for a variety of plausible dust 
mixtures regarding the chemical composition and grain size distributions (for details, see \cite{pantin97}). 
The equilibrium temperatures were found to be in the interval 16\,K to 58\,K 
and depending on the dust albedo and scattering phase function, 
the predicted integrated scattered light, to be consistent with our SEST observations, spans 8 magnitudes.
The two most extreme cases considered were (1) bright cold dust (albedo\,=\,1 at visual wavelengths, 
$T_{\rm dust}=16$\,K) which is
scattering isotropically and has albedo\,=\,0.2 at thermal wavelengths, which results in a spatially integrated
$R$-magnitude of 17.3 and (2) dark warm dust (albedo\,=\,0.02 in the visual, $T_{\rm dust}=58$\,K) which
gives rise to `comet scattering', i.e. 14\% of isotropic at 90\adeg, and has zero albedo at thermal wavelengths,
resulting in an integrated $R$-magnitude of 25.6. These extreme cases are felt to be either too optimistic
or too conservative and an intermediate case might be more appropriate. Our adopted model includes
isotropically scattering dust at $T_{\rm dust}=25$\,K with albedo\,=\,0.2 at visual and zero 
albedo at thermal wavelengths, yielding an integrated $R$-magnitude of 20. An about 10\asec\ source 
($R=25$\,mag/arcsec$^2$) would thus be consistent with both the optical data and our SIMBA measurement
and a deep $R$-band search might become successful. Similarly, the integrated 850\,\um\ flux density is 
predicted to be slightly less than 18\,mJy and should become readily detectable. However, blob {\it C} is
situated outside the figure of \cite{holland98}.

Feature {\it C} is perfectly aligned with the optical disk plane, and its (hypothetical) mass can 
be estimated from 

\begin{equation}
M_{\rm dust} = \frac{D^2\,F_{\nu}}{\kappa_{\nu}\,B_{\nu}(T_{\rm dust})}\,\,\, , 
\end{equation}

where the distance $D$ is assumed to be that of \bpic\ and the adopted {\it dust} absorption 
coefficient $\kappa_{250\,{\rm GHz}} \sim 1$\,\cmg. This estimate of $\kappa_{\nu}$ is 
probably correct within a factor of three (\cite{beckwith2000} and references therein).
For the dust temperature $T_{\rm dust} = 25$\,K, the dust mass is of the order of ten lunar
masses (0.16\,\mearth), which would be comparable to the dust mass of the disk proper, being overall much warmer.
Another factor of about three uncertainty stems thus from the dust temperature, provided
$10\,{\rm K} < T_{\rm dust} < 60$\,K.

The famous `SW-blob' in the 850\,\um\ image has received considerable interest by the debris disk community.
According to \cite{dent2000}, this feature, labelled {\it B} in Fig.\,\ref{bpic_disk}, is real.
It is not readily apparent in our 1200\,\um\ image, however,
presumably due to the combination of low contrast and reduced angular resolution. 
To test this idea, we performed numerical experiments, i.e. convolving two point sources, 
at the appropriate positions of {\it A} and {\it B} and with varying flux ratios, with the SIMBA beam. 
Flux ratios, normalised to the peak value, which are consistent with our observations are in
the range $0.25 - 0.45$, with the most compelling being about 0.3 (comparable to that for blob {\it C}). 
In combination with feature {\it C}, blob {\it B} accounts for the southward bridge seen in Fig.\,\ref{bpic_disk}. 

For radiation mechanisms generating power law spectra and/or thermal dust emission from blob {\it B},
the spectral slope is given by $\alpha = \beta_{A} + 2 - \Delta \log R_{\lambda}/\Delta \log \lambda$,
where $\beta_{A}$, as before, refers to {\it A} = (0\asec, 0\asec), i.e. the \bpic\ disk, and where
$R_{\lambda} = \left [ F_{\nu}(A)/F_{\nu}(B) \right ]_{\lambda}$ and $\lambda = 850$\,\um\
or 1300\,\um. Because of the relatively low S/N of the SCUBA and SIMBA data, the actual flux ratios are 
highly uncertain and, furthermore, calibration uncertainties and telescope beam effects could 
potentially introduce large errors. The combined observations of blob {\it B}
suggest $\log { (R_{850}/R_{1200}) } \sim 0$, yielding $\beta_{B} \sim \beta_{A}$, i.e. consistent 
with the spectrum of {\it A}. 

\section{Conclusions}

Based on 1200\,\um\ imaging observations of the circumstellar dust of \bpic\ we conclude the following:

\begin{itemize}
\item[$\bullet$] At 25\asec\ resolution, the 1200\,\um\ image of the \bpic\ disk is slightly resolved
in the NE-SW direction, but remains unresolved perpendicularly (NW-SE). In addition, the emission appears
asymmetric and extends further SW-S to more than 1000\,AU away from the star. Maximum emission ($=24$\,mJy)
is observed toward the position of the star.
\item[$\bullet$] Combining our 1200\,\um\ map with that by \cite{holland98} at 850\,\um\ we infer that
the dust size distribution in the \bpic\ disk is significantly different from that in the general interstellar
medium and appears more reminiscent of that found in protostellar disks. 
We argue that the thermal emission is dominated by dark big particles and that these grains
constitute a population different from that dominating the scattering in the visible/NIR part of the spectrum.
\item[$\bullet$] From the examination and numerical simulations of the available data (optical and submm/mm)
for the southwestern blobs/extensions seen in the SIMBA maps we conclude that their reality
can at present neither be excluded nor can, on the basis of the available evidence, their existence be 
fully confirmed.
\end{itemize}

\begin{acknowledgements}
We are grateful to Dr. Jos\'e Afonso, who made available to us the Uranus calibration data, and
to the SEST staff for providing additional observations. The critical comments by the anonymous
referee are highly appreciated.
\end{acknowledgements}

\end{document}